# Conceptual and algorithmic development of Pseudo 3D Graphics and Video Content Visualization

**Aladdein M. Amro [1], S. A. Zori [2], Anas M. Al-Oraiqat [*3]**

[1] Taibah University, Department of computer Engineering,
Kingdom of Saudi Arabia
[2] SHEE «Donetsk National Technical University», Ukraine, P.O.Box 85300
[*3] Taibah University, Department of Computer Sciences and Information
Kingdom of Saudi Arabia, P.O. Box 2898

## ABSTRACT

***The article presents a general concept of the organization of pseudo 3D visualization of graphics and video content for 3D visualization systems. The steps of algorithms for solving the problem of synthesis of 3D stereo images based on 2D images are introduced. The features of synthesis organization of standard format of 3D stereo frame are presented. Moreover, the performed experimental simulation for generating complete stereo frames and the results of its time complexity are shown.***

## 1.INTRODUCTION

Pseudo 3D visualization is a generation of visual 3D images on the basis of available images (image based 3D synthesis). This means that a 3D model of the scene is missing or not used for direct 3D image acquisition. The process of pseudo 3D image synthesis is based on the acquisition (recovery, restoration) of a set of projected 3D images of the scene (stereo pairs) in screen space of 3D/stereo display device from previously acquired 2D images of the scene [1],[2]. Such 2D images are generated on the basis of a 3D model as results of shooting and filming etc.

The methods applied in organization of pseudo 3D synthesis use hidden geometry. 3D positions of points are not recovered. But specific algorithms for analysis of images and creating their depth maps are used for conversion of initial images into a stereo 3D image that form the base of 3D restoration, i.e. conversion of initial images into a stereo pair.

It should be noted that known industrial implementations of pseudo 3D synthesis are usually unrevealed closed-technology. This is the case, for example, of synthesis used in modern 3D TVs and screens as a part of such technologies as 3D Hyper Real Engine from Samsung, X-Reality Pro from Sony and Triple XD Engine from LG [3]-[5]. Also such implementations create low-quality images, manifested as g-hosting effects, cross-coupled interferences, insufficient spatial depth of images etc. Moreover, these implementations use a hardware support of special processors, for example, 1.35 GHz quard-core processor ARM Cortex-A15, 1.2 GHz ARM Cortex Dual Core Plus and special multimedia of computing systems of dual-core processor MT5890 as well as quad-core graphical Mali-T624 with 16 GB RAM [3-5].

Manual source frame makeup technologies are used for creation of image depth maps to create high-quality pseudo 3D-images [6],[7]. This technology prevents interactive online generation of pseudo 3D-images. Considering the frequent practical need for pseudo 3D visualization, the development of principles and methods for organization of pseudo stereo 3D graphics /video content visualization and architectures of pseudo stereo 3D visualization computer systems is very critical scientific and practical task.

This paper presents a general concept of the organization of pseudo 3D visualization of graphics and video content for 3D visualization systems. In addition, an algorithm for solving the problem of synthesis of 3D stereo images based on 2D images is proposed. Experimental simulation of generating complete stereo frames and the results of its time complexity are also given.

The rest of this paper is organized as follows: First, Section 2 presents mechanisms and algorithms for Pseudo 3D stereo/2D images acquisition. Next, Section 3 introduces the proposed inpainting algorithm. Then, Section 4 details the generation process of a standard stereo frame from a stereo pair. Section 5 experimentally presents and analyzes the proposed 2D images conversion into 3D pseudo stereo pair. Finally, Section 6 gives the conclusion.

## 2.PSEUDO 3D STEREO/2D IMAGES ACQUISITION: MECHANISMS AND ALGORITHMS

Considering the above literature review and the closeness of most existing technologies for “intellectual” pseudo 3D-images creation, it is proposed to realize the process for acquisition of pseudo stereo 3D images from 2D images on the basis of the scheme shown in Figure 1.

A recent 2D-to-3D conversion system using the edge information method of [8] is taken as basis since it gives good practical results and offers block-based frame processing algorithms. Accordingly, enhancing generation process through parallelizing is allowed.

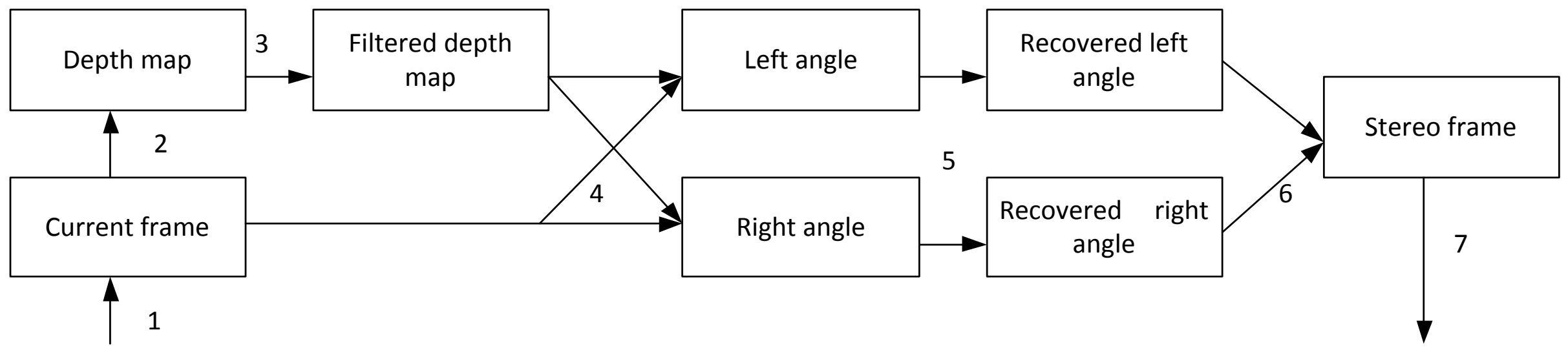

**Figure 1** Proposed scheme of generation of pseudo stereo 3D image from 2D image.

The main processes of the proposed scheme are as follows:
1) Source image acquisition (probably, extraction of the frame from media stream).
2) Acquisition of depth map of the current frame.
3) Depth map filtration (by cross-bilateral filtering).
4) Creation of stereo pair from the current frame and filtered depth map.
5) Restoration of angles, considering possible defects (filling of gaps in acquired angles - Inpainting).
6) Conversion of a stereo pair into a standard format of 3D frame.
7) Possibly adding the generated 3D frame to the resultant 3D video stream.

It should be noted that Stages 1, 6, and 7 are compulsory. The 1st and 7th stages are used only for handling video streams. As for the 6th stage, it can be realized by 3D video device hardware. Thus, the proposed generalized algorithm for generating a 3D stereo frame from 2D images is presented in Figure 2.

```
Begin
  Open_image;
  Generate_depth_map;
  Filter_depth_map;
  Generate_ stereo pair;
  Apply _Inpainting_for _left_angle;
  Apply_Inpainting_for right_angle;
  Generate _stereo frame;
  Save _acquired _image;
End
```

**Figure 2** Proposed algorithm for generating pseudo stereo 3D image from 2D image.

The details of each stage of the proposed algorithm are explained as follows:
**Stage I:** Algorithms for extracting the image objects and creating a depth map.
To generate a depth map it is necessary to extract the image objects. The depth of these objects (i.e. the distance from the observer) will be further used for creating image depth map.
The process of extraction of image objects is one of the classical data processing tasks. The following algorithms are most popular [2]:
- Algorithms directly extract object shapes from the image;
- Algorithms create objects in the form of pixel region arrays.

Image shape extraction algorithms are based on a basic property of brightness signal, known as discontinuity. Spatial filtration by masks is used for this purpose. Sobel's, Prewitt's and Robert's are among the well-known masks [9,10]. Algorithms of this group give good results. But they are computationally complex since they require calculation of the 1st and 2nd order derivatives for each direction and joint gradient values.

Thus, discrete analogues of the 1st and 2nd order derivatives and also gradient are used for detection of brightness abrupt changes [9],[10]. The selection of mask type and size is unspecified and has a significant effect on execution speed of the algorithms of this class and the quality of the filtered object shapes. Therefore new Rost's region (growing) algorithm is introduced a representative of the second class algorithms for extraction of the image objects [10]. Region growing algorithms create objects in the form of pixel region arrays. According to literature, this algorithm is quick with high-quality.
By analyzing the brightness of each pixel of the image, the idea of this algorithm is to mark the image region pixels that have some degree of similarity in color (brightness) as similar.

The algorithm can be described as follows:

1. Upper left pixel of the image is a new class $\mathbf{C_1}$.
2. For the pixels of the first image line, the deviation $\boldsymbol{g}$ from the class of the left neighbor pixel is computed and compared with a certain threshold $\boldsymbol{\delta}$. If the deviation from the class is less than the threshold, the pixel is inserted into the class of its neighbor, otherwise a new class is created.
3. The first class of each further line is compared with the class of the first pixel. Then current pixel is compared with classes of two neighbors, i.e. left and upper pixel. For each further line the current pixel is compared with the classes of the two left and upper neighbors (pixels). In this case:
   3.1 If the deviations from both classes exceed the threshold, a new class is created. If the deviation is more than only one class, the pixel is inserted into the other class having deviation less than the threshold.
   3.2 If the deviation is less than the threshold for both classes, there are two possibilities:

   A) $|\boldsymbol{g}(C_{Left}) - \boldsymbol{g}(C_{Upper})| \le \boldsymbol{\delta}$ then these 2 classes are combined and the current pixel is added to the combined class.
   B) $|\boldsymbol{g}(C_{Left}) - \boldsymbol{g}(C_{Upper})| > \boldsymbol{\delta}$ then the pixel is added to the class with the minimal deviation.

Once the objects are extracted for the current image, it is necessary to determine the distance of each object from the observer. Depth map generation algorithms are used for solution of such task. The depth map is a black and white image, on which object depth is determined by their brightness gradient (the closer, the brighter).
This task is nontrivial since it is hard (and impossible sometimes) to accurately determine the remoteness of each object from the observer on the basis of a 2D image. Therefore, there is no correct and accurate solution of this task. On the other hand, it is a very significant stage of creation of pseudo 3D images, the quality of which directly affects the quality of acquitted 3D stereo frames.
In literature, there are several approaches to the creation of the depth map for a 2D image [11]:

- Depth map based on depth effects from landscape scene.
- Depth map based on depth from motion.
- Depth map based on depth from focus.
- Depth map based on depth from geometry.

Based on analysis of these existing methods for obtaining depth map from 2D images, it was decided to use *depth effects from landscape scene* method as the most universal method with less computation. However, the *depth from motion* method requires the use of video files and object motion data, *depth from focus* method requires time-consuming calculations and restoration of camera position and *depth from geometry* method offers good results only upon processing of scenes where straight lines and geometries prevail.
The *depth effects from landscape scene* algorithm are based on the idea that the objects located in front bottom position of the image are closer to the observer than the objects in upper part (spatial perspective effect). Basing on this fact, it is possible to make conclusion with respect to remoteness of the object from the observer. Scene depth is represented by vertical brightness gradient of the image from white to black.
Considering this fact and the use of previously described region extraction algorithms, depth map generation algorithm can be expressed as follows (Figure 3):

```
Begin
   Do (i=0 to i< number_of_areas )
   Begin
     area_height:=0;
     Do (j=0 to j< number_of_pixels_in_the_area)
     Begin
       If area_region >height_of_current_area_pixel
       Then
```

**Figure 3** Depth map generation algorithm.

```
        Begin
          area_height:=height_of _current _area_pixel;
        End
      End
      Do (j=0 to j< number_ of_pixels_in_the_area )
      Begin
        Pixel_depth:=area_height/image_height;
      End
    End
End
```

**Figure 3** (Continued).

The pixel depth value of the current region is calculated by a simplified equation in the proposed algorithm (Figure 3). This formula requires considerably less calculations. It also ensures higher performance and almost identical quality of image depth map (determined experimentally for test scenes). The resultant depth will be assigned to all pixels of a region. Its distribution among the objects is planned to be corrected further by joint usage of the obtained and filtered depth map (depth interpolation).

To remove artifacts and get final distribution (application) of the depth map, it is necessary to carry out cross-bilateral filtration of the map [11]. Objects (regions) on resultant depth map will remain 'flat' without filtration since they have the same depth for all pixels.

A depth map filtration algorithm was developed on the basis of [11] and the following equations 1 – 3.

$$D_f = \frac{A_i}{N(x_i)} \quad (1)$$

$$A_i = \sum_{x_j \epsilon \Omega} D_{x_j} * e^{\frac{\sqrt{(x(x_j)-x(x_i))^2+(y(x_j)-y(x_i))^2}}{2\sigma_S^2} - \frac{\sqrt{(I(x_j)-I(x_i))^2}}{2\sigma_C^2}} \quad (2)$$

$$N(x_i) = \sum_{x_j \epsilon \Omega} e^{\frac{\sqrt{(x(x_j)-x(x_i))^2+(y(x_j)-y(x_i))^2}}{2\sigma_S^2} - \frac{\sqrt{(I(x_j)-I(x_i))^2}}{2\sigma_C^2}} \quad (3)$$

where:

$D$ is the value of the current pixel depth.
$\sigma_s$ is a parameter of cross-bilateral filter.
$\sigma_c$ is a parameter of cross-bilateral filter.
$\Omega$ is a subset of core pixel.
$x_j$ is the current core pixel.
$x_i$ is the current pixel.
$I(x_j)$ is the brightness of the current core pixel.
$I(x_i)$ is the brightness of the current pixel.
$x(x_j)$ is the X-coordinate of the current core pixel.
$y(x_j)$ is the Y-coordinate of the current core pixel.
$x(x_i)$ is the X-coordinate of the current pixel.
$y(x_i)$ is the Y-coordinate of the current pixel.

The filtration upon distribution of depth among objects in the resultant image can cause distortion of the depth map. This is due to the application of the proposed simplified model of determining the depth of pixel regions (Figure 3). Hence, for more accurate depth of pixels, corrections of the results of the map, filtered by equations (1-3), is performed with the initial depth map. The depth of object pixels is represented by black/white gradient (0-255).

**Stage II:** Algorithm for creating stereo pair from an initial image and depth map.

A stereo pair can be created on the basis of a generated depth map and an initial image. When generating the left and right view, stereoscopic parallax uses the concept of [12]. According to this concept, the visual object position changes depending on the point of observation. In accordance with [13]-[15], the position of the object against the screen is considered upon generation of stereo pair. If the object is after some virtual screen plane, it is called object with a positive parallax. If the object is before the virtual screen plane, it is called a negative parallax (Figure 4).

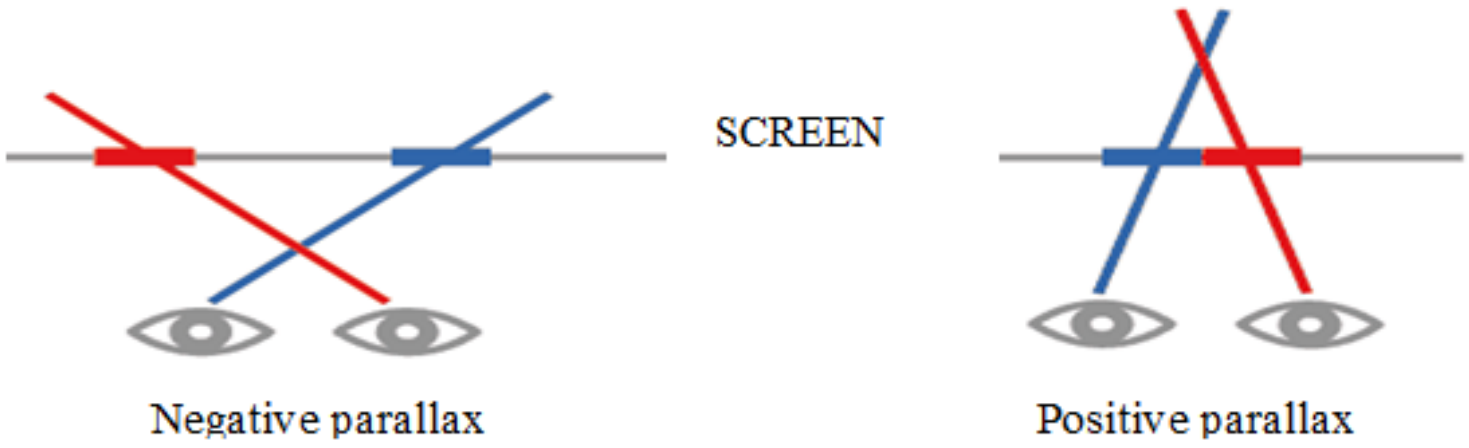

**Figure 4** Stereoscopic parallax.

As illustrated in Figure 4, the larger distance of the object from the screen, the higher value of its displacement for each eye. So, a human perceives objects that are located before the screen as obviously volumetric. The closer they are to virtual point, the higher 3D effect is.

Thus, the general idea of the algorithm for generating a pseudo stereo image is as follows: It is necessary to select the position of a virtual screen plane and calculate new coordinates of pixels of the objects from an initial image (horizontal displacement) for the left and the right perspectives, depending on the position of the object against the virtual screen plane (set by the resultant depth map).

Objects, which are close to white, are closer to the observer in the resultant depth map and are placed before the virtual screen. Therefore, the formulas (4 and 5) for calculating the coordinates of the current pixel for the left and the right perspectives of stereo pair are represented below:

$$x_l = \begin{cases} x + \frac{B}{2} * \frac{d_x}{255}, if\ d_x > 150 \\ x - \frac{B}{2} * \left(1 - \frac{d_x}{255}\right) or \end{cases} \quad (4)$$

$$x_r = \begin{cases} x - \frac{B}{2} * \frac{d_x}{255}\ if\ d_x > 150 \\ x + \frac{B}{2} * \left(1 - \frac{d_x}{255}\right)\ or \end{cases} \quad (5)$$

where:
$x_l$ is the x-coordinate of the current pixel for the left perspective ;
$x_r$ is the x-coordinate of the current pixel for the right perspective;
x is the x-coordinate of the current pixel;
$d_x$ is the depth value of the current pixel;
B is the basis value (displacement for cursors).

The value "150" is associated with the position of the virtual screen plane (0 - 255) and it has been found by experience as giving an acceptable result of volume, but it can be changed. It should be noted that this value can also vary and determined based on the desired '3D effect' ('weak', 'normal' and 'strong') in existing practical implementations of similar systems, however too small values cause artifacts and visual discomfort. The proposed stereo pair generation algorithm for pseudo 3D is shown on Figure 5.

```
Begin
  Do (i=0 toi=frame_width)
  Begin
    Do (j=0 to j=frame_height)
    Begin
      If Pixel _depth (i,j)>150
     Then
        Begin
          x_coordinate_of_left_frame:= i +Basis/2*Pixel_depth(i,j) / 255;
          x_coordinate_of_left_frame:= i - Basis/2*Pixel_depth(i,j) / 255;
        End
     Else
```

**Figure 5** Pseudo 3D-stereo pair generation algorithm.

```
            Begin
                x_coordinate_of_left _frame:= i- Basis/2*(1- Pixel_depth(i,j) /255);
                x_coordinate_of_right _frame:= i+ Basis/2*(1- Pixel_depth(i,j) /255);
              End
            If x_coordinate_of_left_frame > 0  And  x_coordinate_of_left _frame < frame_width
            Then
              Begin
                Set_pixel_value_in_left_perspective (x_coordinate_of_left_frame,j, Initial_color);
              End
            If x_coordinate_of_right _frame >0 And x_coordinate_of_right_frame <frame_width
            Then
              Begin
                Set_pixel_value_in_right_perspective(x_coordinate_of_right _frame,j,Initial_color);
              End
          End
        End
      End
```

**Figure 5** (Continued).

## 3.INPAINTING ALGORITHM

The creation of stereo pair may result in 'gaps' with missing a part of the initial image. Therefore it is necessary to recover the lost information for perspectives. The problem associated with insufficient information is a classical image processing problem. Algorithms that solve this problem are called inpainting algorithms.

There are different approaches to solve this problem [16]-[18]. It should be noted that there are no detailed implementation description of inpainting algorithms and comparisons of their efficiency. Thus, the developed inpainting algorithm is mainly founded on the basis of the exemplar based inpainting algorithm [18]. The ideal of the algorithm is based on gradual dissemination of the information along the damaged area from its points bordering undamaged pixels to the points not bordering undamaged pixels. Thus the functions of the developed inpainting algorithm are as follows:

1. Damaged pixels are organized into areas.
2. The area recovery starts from pixels bordering undamaged pixels.
3. The average color value of the current damaged pixel is calculated as the mean value of its connected pixels.

In order to speed up the algorithm, it is realized as an incremental algorithm.

Pseudo code of the inpainting algorithm is shown in Figure 6.

```
Begin
  Initialization of pixel damage list ();
  Initialization of matrix of all image pixels ();
  Do (i=0 toi=Frame _width)
  Begin
    Do (j=0 toj=Frame_height)
    Begin
      If current _pixel (i,j) is damaged
      Then
      Begin
          Add current_pixel (i,j) to the list of_damaged_pixels;
      End
    End
    Number_of_damaged_pixels:= Number_of _elements In list of_ damaged_pixels;
    While number of _damaged_ pixels!=0
    Begin
       Do (i=0 to i=Number_of_elements_list_of_damaged_pixels)
```

**Figure 6** Inpainting algorithm.

```
        Begin
          If current_pixel (i,j) is damaged
          Then
          Begin
            Get_number_of_damaged_neighbors (current_pixel (i,j));
            If number_of_undamaged_neigbours >=2
            Then
            Begin
              Change current_pixel(i,j) by NOT_damaged;
              Number_of_damaged_pixels--;
              Calculate_color_of_current_pixel(i,j);
              Set_value_of_current_pixel_of_result_representation(i,j,Pixel_color(i,j));
            End
          End
        End
      End
    End
  End
End
```

**Figure 6** (Continued).

## 4. GENERATION OF STANDARD A STEREO FRAME FROM A STEREO PAIR

Once a stereo pair is created from the current frame it is necessary to create a stereo frame of standard 3D-stereo format. It should be mentioned that the execution of this procedure is not compulsory. It is used for creating the structure of 3D video stream in one of the standard presentation formats and also for preparing to output stereo 3D-images matching standard 3D display devices.

The following algorithmic restructuring of the obtained stereo pair are executed depending on what standard 3D stereo format is required:

1. Frame-result should be twice larger than current resources in the horizontal direction. Also simple composition is performed for a full-size horizontal stereo pair.
2. Frame-result should be twice larger than current resources in the vertical direction. Also simple composition is performed for full-size vertical stereo pair.
3. It is necessary to compress the left and right perspectives twofold in the horizontal direction in comparison with the original image. Moreover, it is necessary to record the compressed left perspective into the left part of frame-results and the right perspective into the right part of frame-results to receive an anamorph horizontal side-by-side stereo pair (and vice versa for a crossed stereo pair).
4. It is necessary to compress the left and right perspectives twofold in the vertical direction in comparison with the original image. Moreover, it is necessary to record the compressed left perspective into the left part of frame-results and the right perspective into the right part of frame-results to receive an anamorph vertical stereo pair (and vice versa for crossed stereo pair).
5. The frame-result forms an image where lines from the left and right frames alternate for interlaced format. In this case, the initial frames are compressed twofold or only even/odd frames are selected from the initial frames.
6. The value of each pixel of the stereo frame is calculated on the basis of information about colors of the given pixel in each perspective for the anaglyph format [19].

It should be noted that the anaglyph format is one of the oldest formats that has color rendering issues. But it is the most universal since it allows reproducing 3D-images on non-3D display units. On the other hand, other formats require 3D display units for reproduction.

## 5. EXPERIMENTAL STUDY OF 2D IMAGES CONVERSION INTO 3D PSEUDO STEREO PAIR

Experimental study of the proposed algorithms for 2D images conversion into 3D pseudo stereo pair has been done. The test was carried out for two test beds (Table 1).

**Table 1**: Configuration of test beds.

| OS | CPU | RAM |
|---|---|---|
| Windows 7 Ultimate, x86 | Intel Pentium Dual Core T4300 @ 2.1 GHz | 4 Gb, DDR3 |
| | Intel Core 2 Quad Q9550 @ 2.83 GHz | |

The experiment was carried out on the frame taken from Star Wars: Episode II film [20]. Figure 7 shows the experimental results for the time of standard 3D stereo pairs full generation (anaglyph and horizontal anamorph) from 2D-image. Initial images have been set with resolution of 320x480, 800x600 and 1920x1080 pixels. The results are without reference to the disk access time for reading and recording the initial and resultant files.
The obtained images as results of execution of this algorithm are shown on Figure 8. The experimental results show that the proposed approach and algorithms for generation of 3D pseudo stereo images from 2D-images are reasonable but online generation of pseudo stereo 3D -images for high resolution frames is not feasible. Therefore it is necessary to study the architecture of the hardware support.

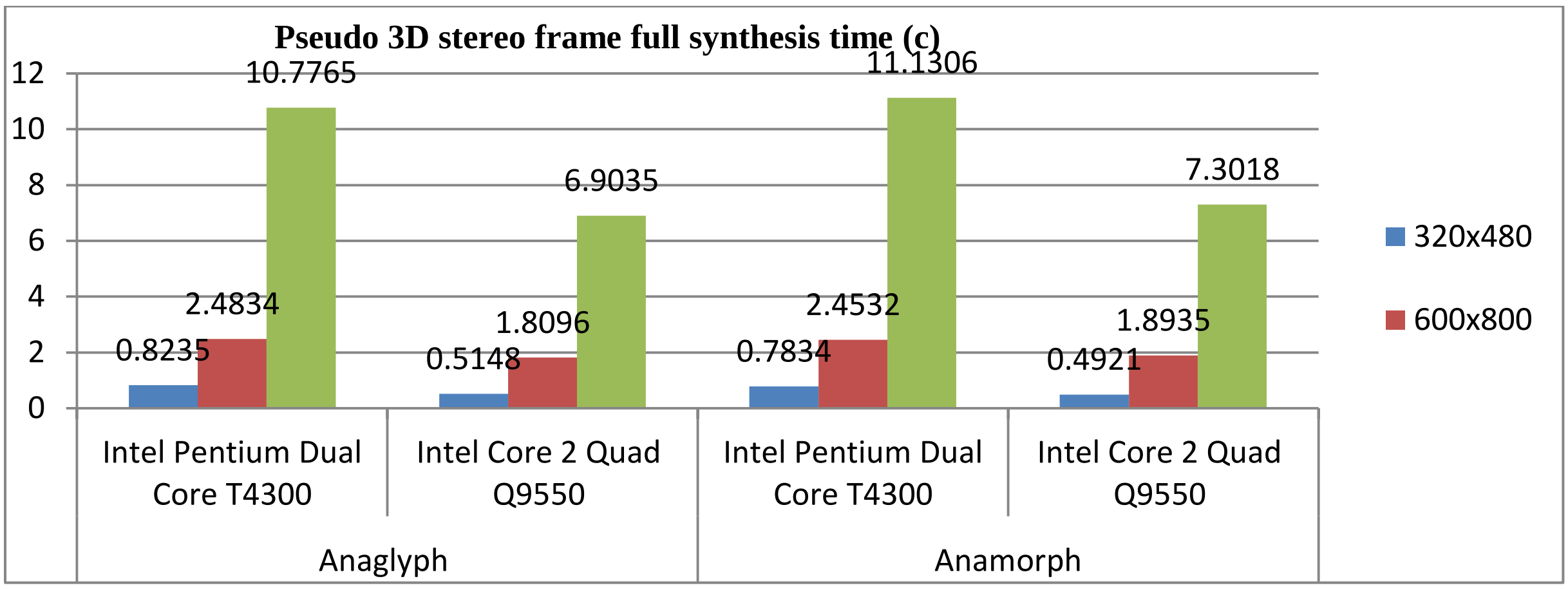

**Figure 7** Time of full synthesis of pseudo 3D image from 2D image.

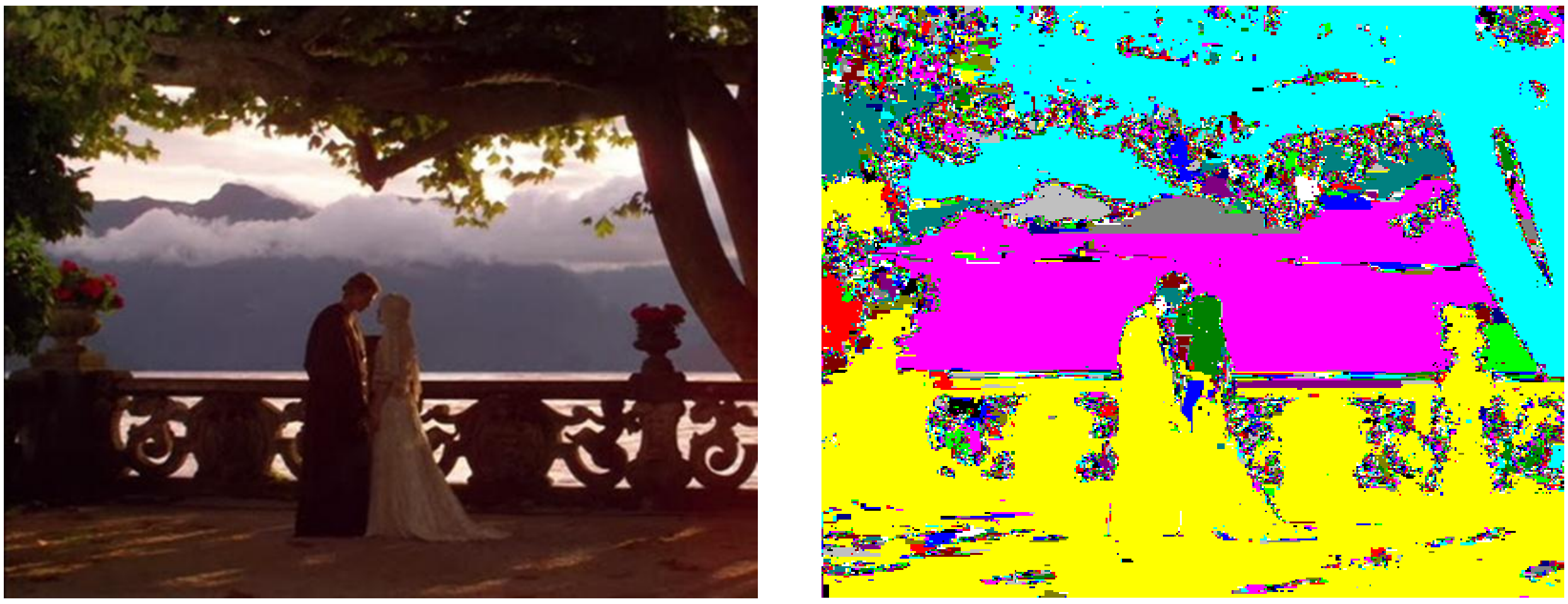

a) Initial 2D frame.

b) Extracted areas.

**Figure 8** Results of experimental generation of 3D pseudo stereo image from 2D image: a) Initial 2D frame. b) Extracted areas. c) Primary depth map. d) Filtered resultant depth map. e) Generated left perspective. f) Generated right perspective. g) Left perspective after inpainting. h) Right perspective after inpainting. i) Stereo frame in Color Anaglyphs format. j) Stereo frame in Anamorph Horizontal format.

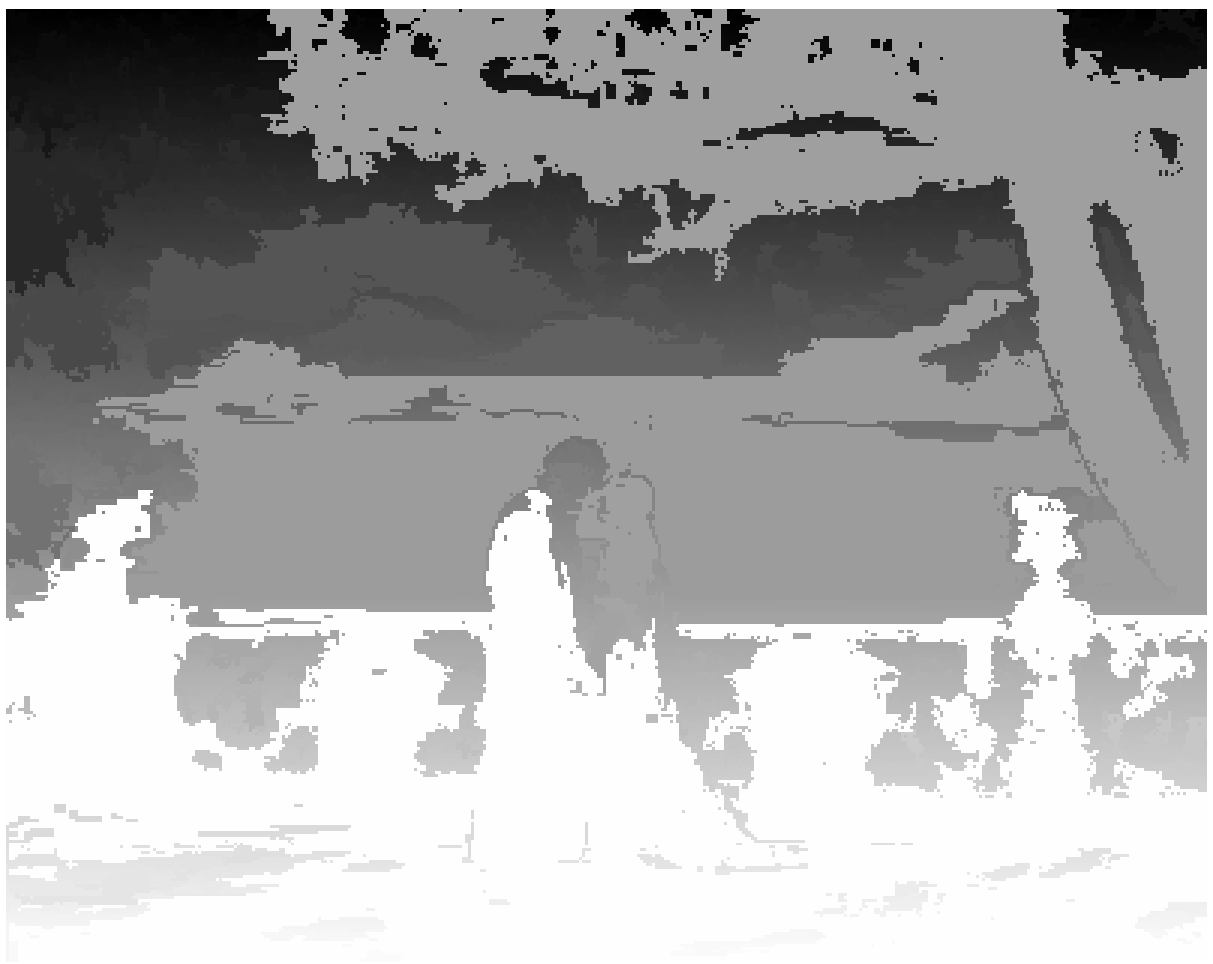

c) Primary depth map.

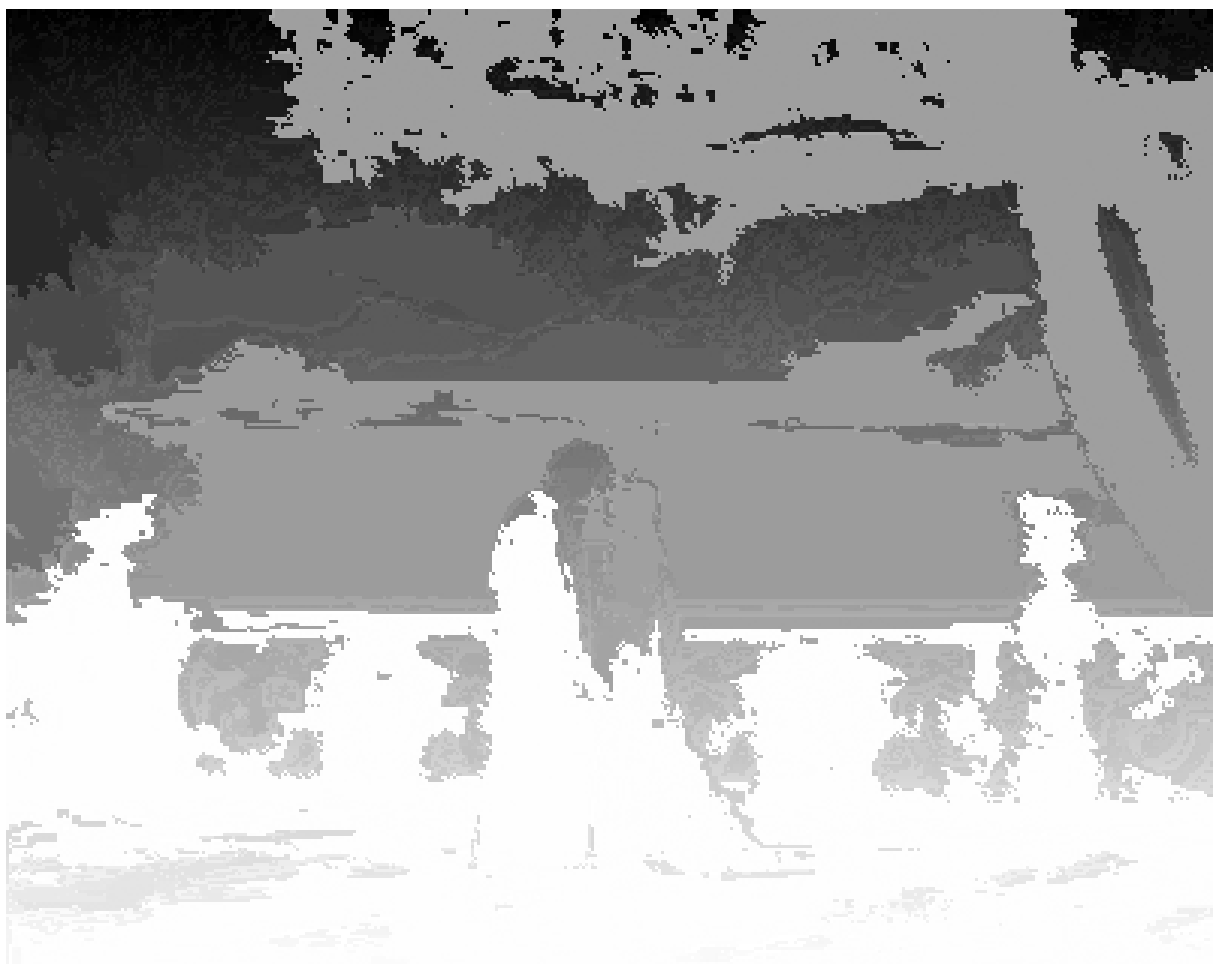

d) Filtered resultant depth map.

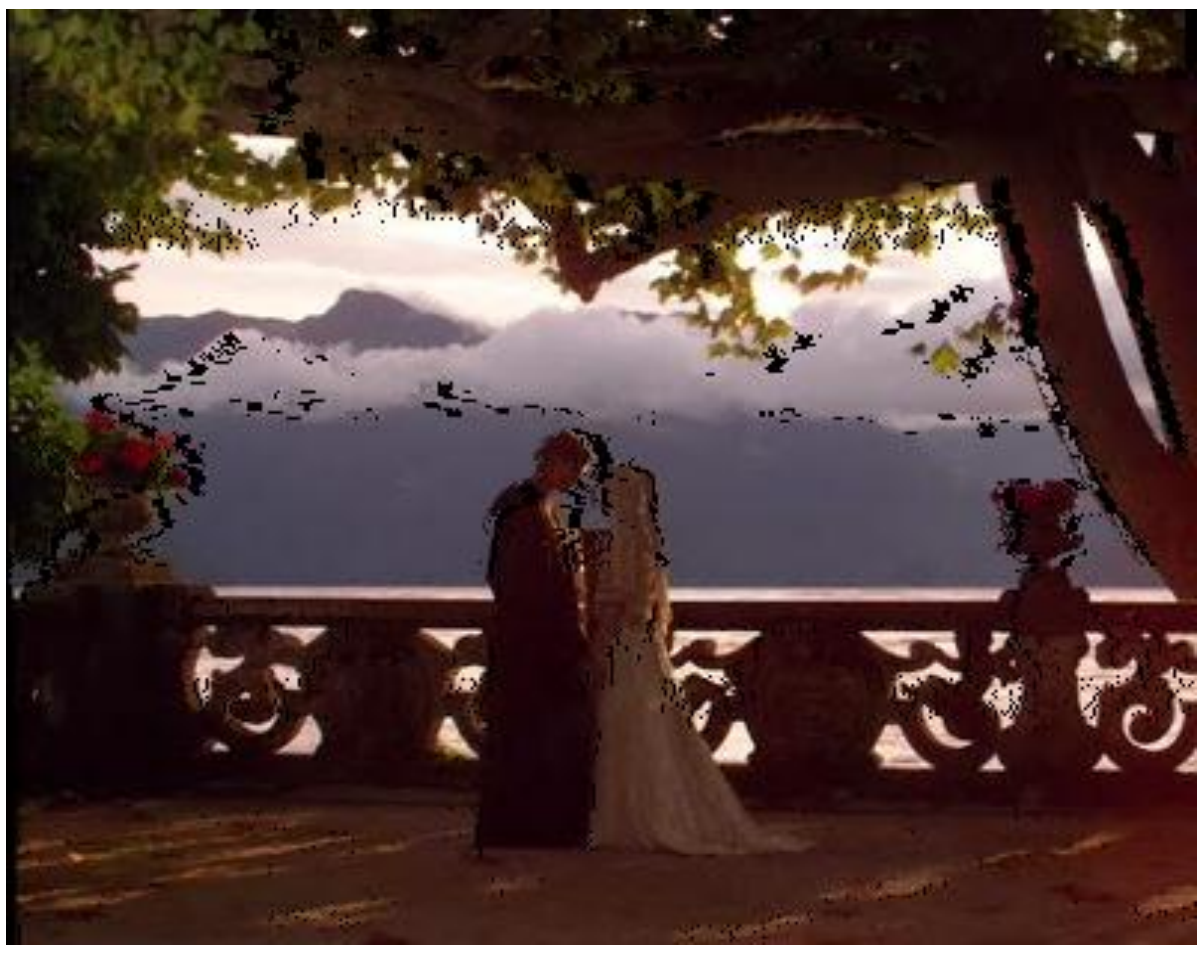

e) Generated left perspective.

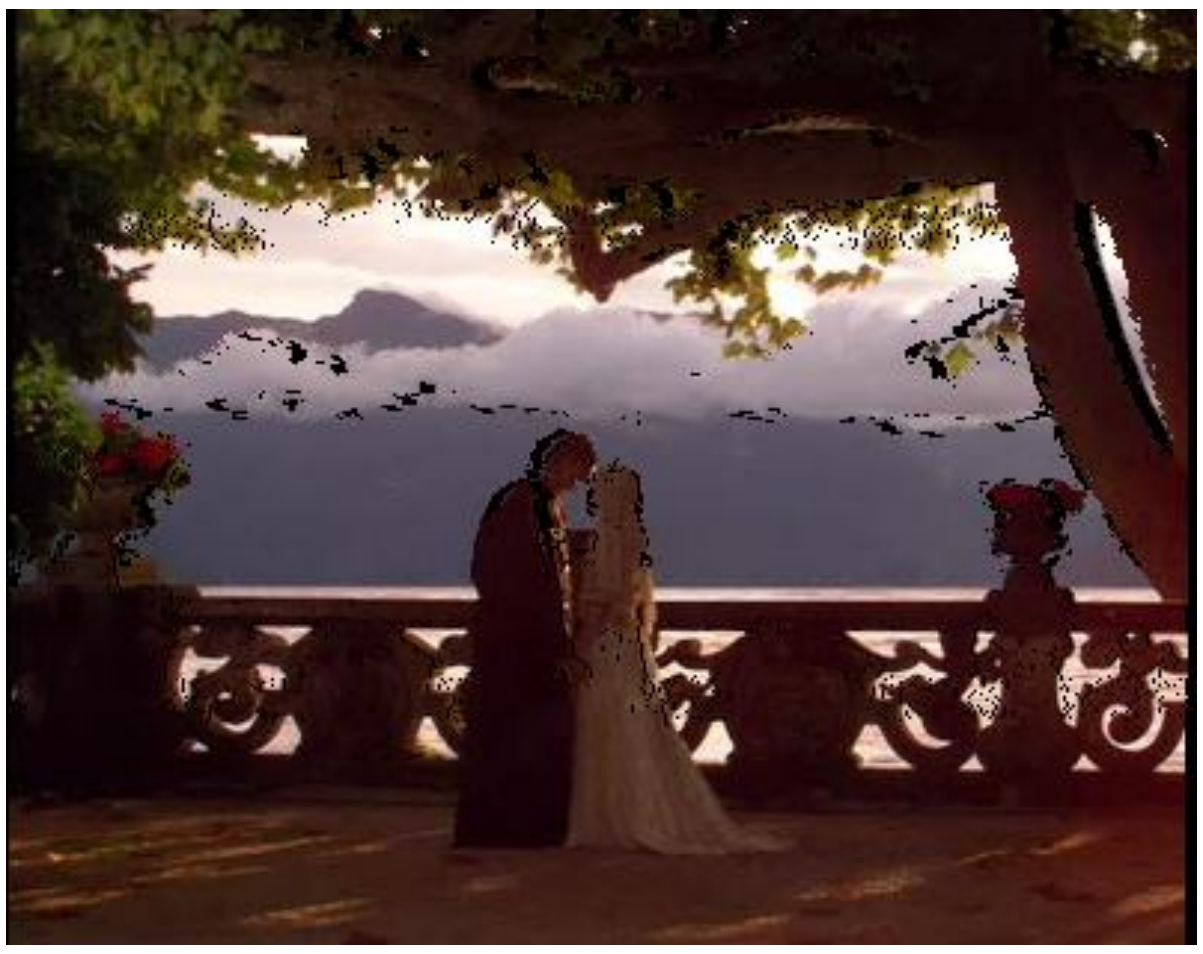

f) Generated right perspective.

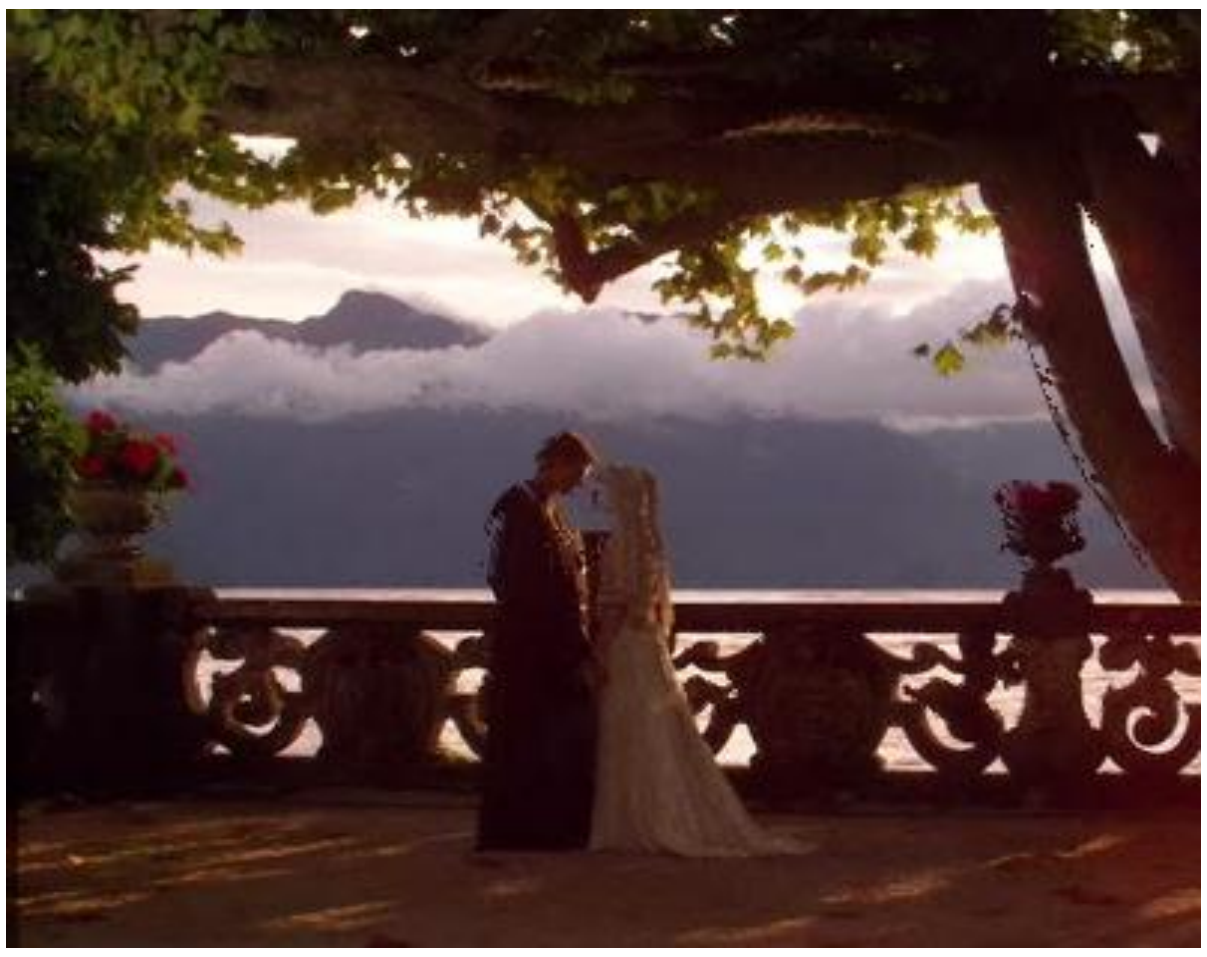

g) Left perspective after inpainting.

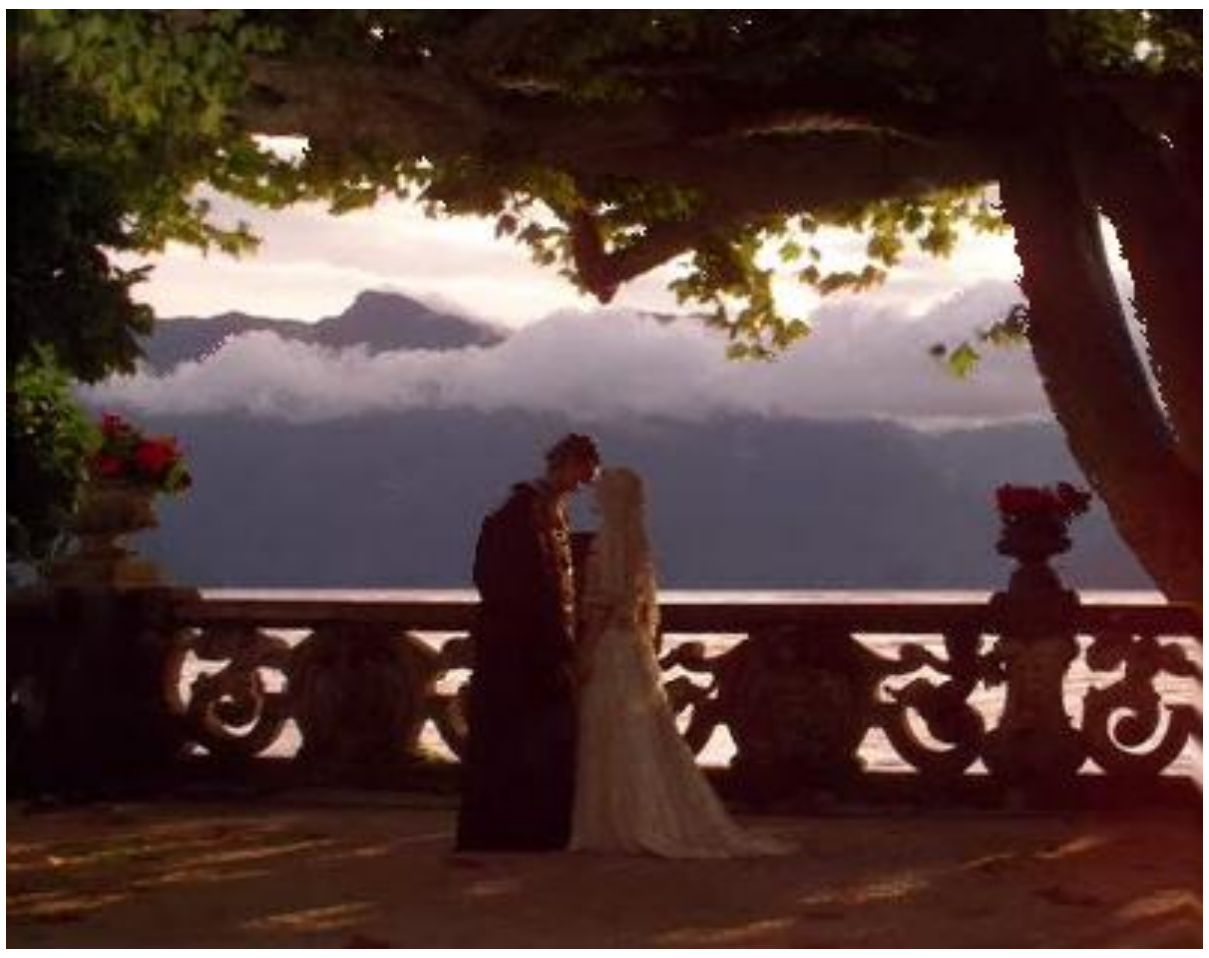

h) Right perspective after inpainting.

**Figure 8** (Continued).

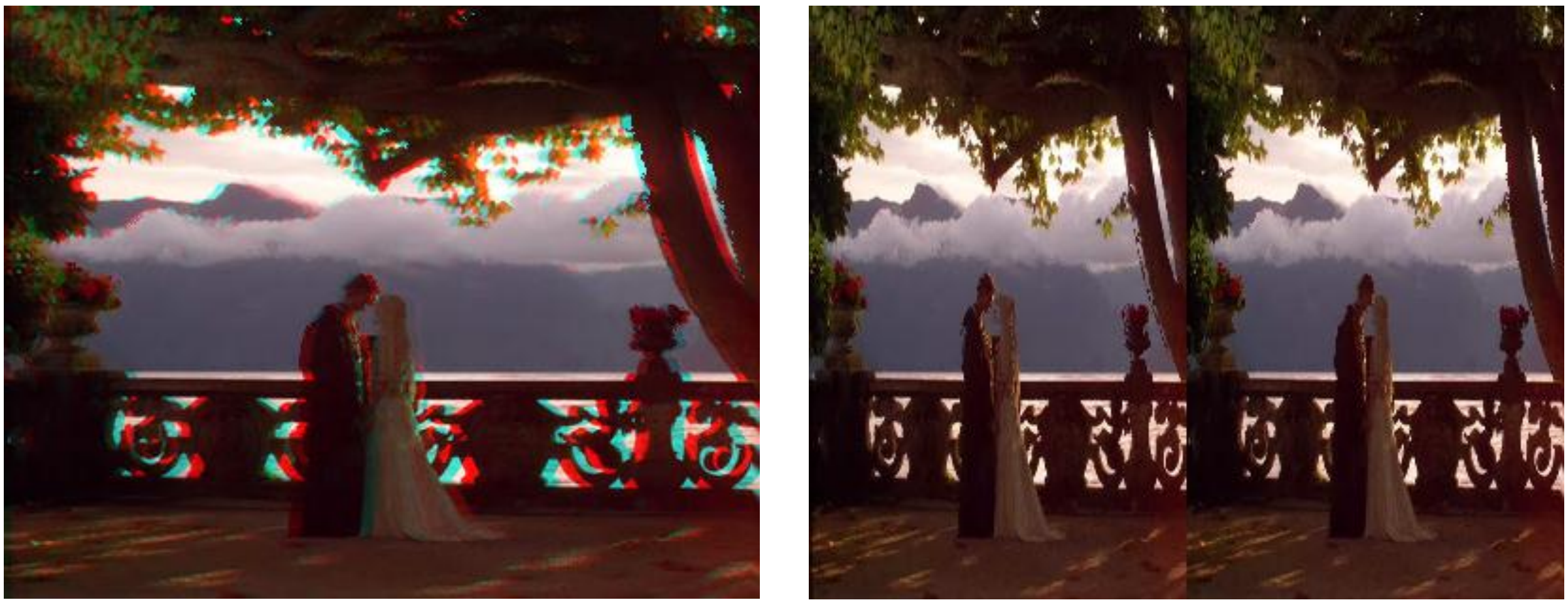

i) Stereo frame in Color Anaglyphs format. j) Stereo frame in Anamorph Horizontal format.

**Figure 8** (Continued).

The organization of parallel hardware support resources, for the developed algorithms on the basis of General-Purpose GPU-CUDA architecture, is offered to improve the performance of the system of realistic 3D pseudo stereo visualization. The possibility for realizing each stage of 3D stereo pseudo synthesis on the basis of this architecture has been analyzed. It has been shown that it is not efficient to translate the depth map into CUD-syntax. A simplified solution, without significantly affecting the quality, is offered for realizing an integrated Inpainting algorithm. Hence, it allows realizing CUDA of this algorithm. CUDA-realization of process stages has been performed. In addition, the efficiency of the offered algorithmic base realization, for 3D pseudo stereo image synthesis on two GPU test systems, was studied. The results of the carried out experiments are illustrated in Figures 9 and 10. Based on these results, the following points can be noted:

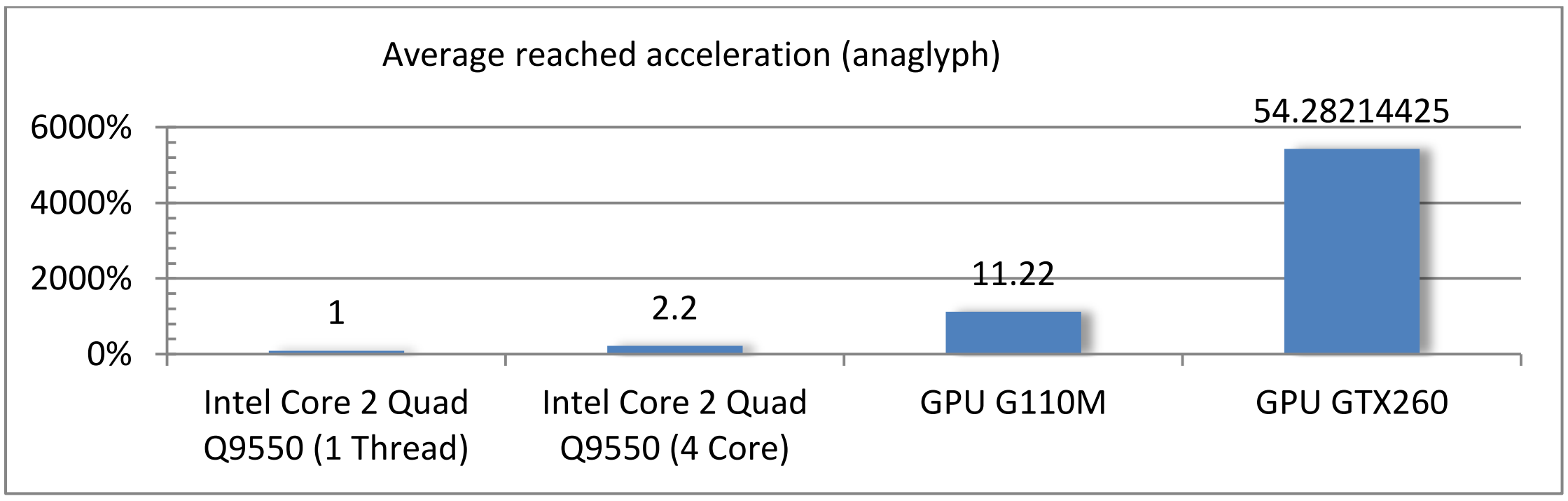

**Figure 9** Acceptable acceleration of pseudo stereo 3D synthesis (anaglyph).

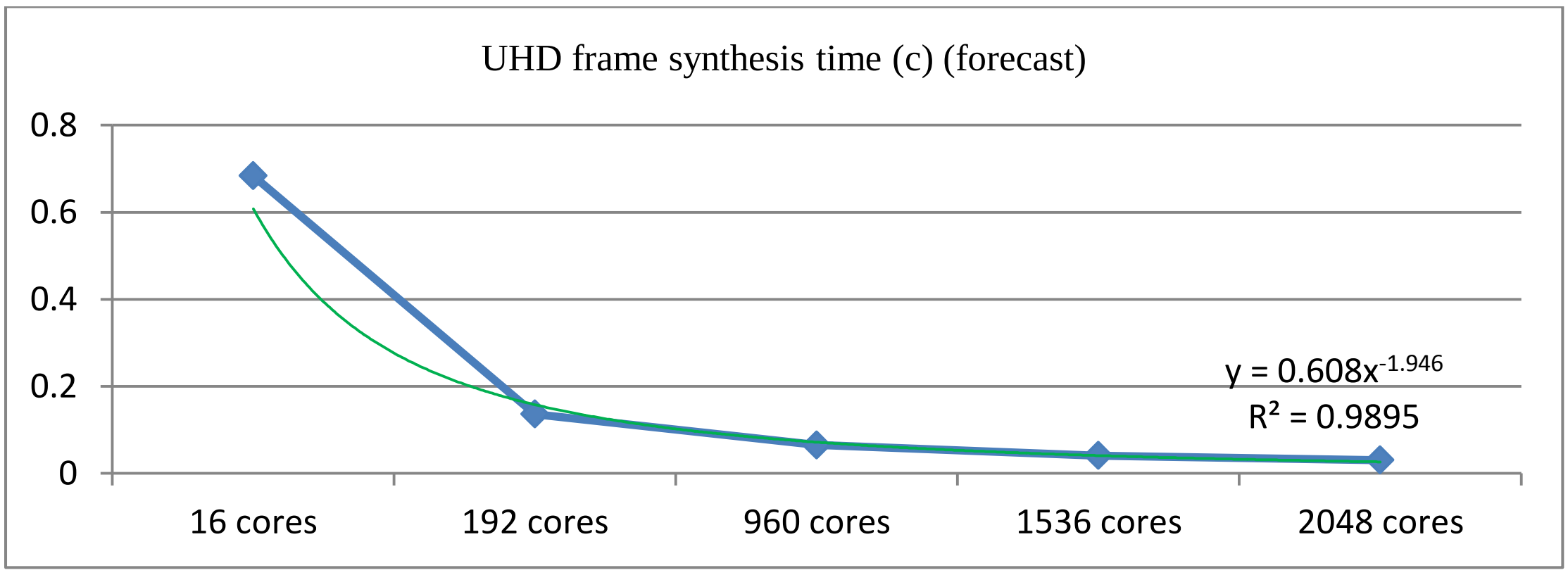

**Figure 10** Forecast of performance growth upon realization on GPU systems.

- The offered generation adaptation of 3D pseudo stereo images, to the parallel software support architecture of GPU systems is efficient. It also considerably improves the performance except for the depth map generation stage that has serial processing mechanism.
- For the test GPU, acceptable computation procedures acceleration of 3D pseudo stereo synthesis for anaglyph and anamorphic 3D stereo frame formats is equal to 11 and 54 times on average without execution of optimized procedures.
- The sizes variations of CUDA network blocks give additional acceleration for test GPU by 1.25 times on the average. This fact confirms the relevancy of correct selection of the computation network configuration and other optimizations.
- Without considering critical characteristics of GPU such as type of microstructure, GPU operation frequency, memory etc., theoretical prediction of synthesis performance improvement with GPU core quantity shows that the realization of modern GPU system (more than 2000 cores) allows online realization of the process.

## 6.CONCLUSION

This paper presents a general concept of pseudo 3D-visualization of graphical and video content organization for 3D-visualization systems. The basic needed stages are determined and hence algorithms for obtaining solution of 3D-stereo image problems on the basis of 2D-images are realized.
Also the paper demonstrates main features of standard stereo 3D-image synthesis organization, experimental modeling of full stereo frame generation process and evaluation of its time complexity. Experiments results show that the proposed approach and algorithms for generation of pseudo stereo 3D-images from 2D-images are efficient but online generation of pseudo stereo 3D-images for high resolution frames is not feasible.
Moreover the paper offers adaptation of pseudo stereo 3D-images to the architecture of parallel hardware support of GPU-system. This adaptation is efficient and results in considerable increase of the efficiency except for the depth map generation stage that has a serial processing mechanism. The realization of the system of modern GPU system ( more than 2000 cores) allows online realization of the process.

## ACKNOWLEDGMENT

The authors would like to thank Taibah University and Donetsk National Technical University for supporting this research. Also, the valuable comments of the anonymous reviewers are greatly appreciated.

**Author's Profile**

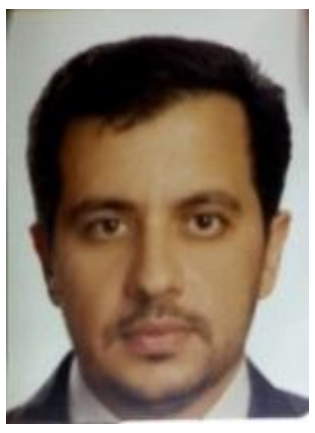

**Aladdein M. Amro** received M.S. in Automation Engineering from Moscow Technical University in1996, and Ph.D. in Telecommunications Engineering from Kazan State University (Russian Federation) in 2003. Had been an Assistant Professor at the Computer Engineering Dept. Al-Hussein Bin Talal University (Jordan) during the years 2004-2011. Since then has been working as an Assistant Professor at the Computer Engineering Dept., Taibah University (Kingdom of Saudi Arabia). Research interest is in the areas of digital Signal processing, image processing, real time systems.
E-mail: amroru@hotmail.com

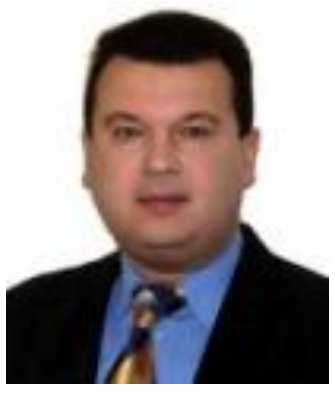

**Sergii A. Zori** received a Ph.D. in Computer Systems & Components from NTU "Donetsk National Technical University" (Ukraine) in 1996 and Associated Professor in 2017. He is Associate Professor at the Mathematics & Computer Science Department of the Faculty of Computer Science and Technologies, Donetsk National Technical University ("DonNTU").
His research interest is in the areas of computer graphics, image/video processing & vision, 3D devices, Virtual reality systems, specialized parallel computer systems.
E-mail: sa.zori1968@gmail.com

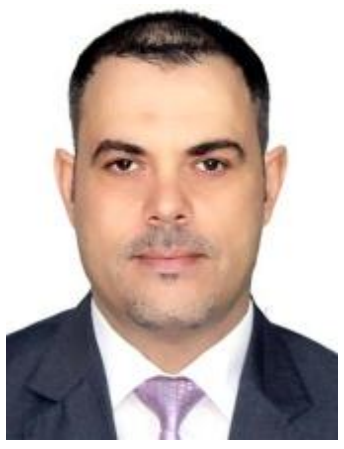

**Anas Mahmoud Al-Oraiqat** received a B.S. in Computer Engineering and M.S. in Computer Systems & Networks from Kirovograd Technology University in 2003 and 2004, respectively, and Ph.D. in Computer Systems & Components from Donetsk National Technical University (Ukraine) in 2011. He has been an Assistant Professor at the Computer & Information Sciences Dept., Taibah University (Kingdom of Saudi Arabia) since Aug. 2012. Prior to his academic career, he was a Network Manager at the Arab Bank (Jordan), 2011-2012. Also, he was a Computer Networks Trainer at Khwarizmi College (Jordan), 2005-2007.
His research interest is in the areas of computer graphics, image/video processing, 3D devices, modelling and simulation of dynamic systems, and simulation of parallel systems.
E-mail: anas_oraiqat@hotmail.com